\documentclass[12pt]{article}

\usepackage{cite}  

\begin{document}

\def\CA{{\cal A}}
\def\CB{{\cal B}}
\def\CC{{\cal C}}
\def\CD{{\cal D}}
\def\CE{{\cal E}}
\def\CF{{\cal F}}
\def\CG{{\cal G}}
\def\CH{{\cal H}}
\def\CI{{\cal I}}
\def\CJ{{\cal J}}
\def\CK{{\cal K}}
\def\CL{{\cal L}}
\def\CM{{\cal M}}
\def\CN{{\cal N}}
\def\CO{{\cal O}}
\def\CP{{\cal P}}
\def\CQ{{\cal Q}}
\def\CR{{\cal R}}
\def\CS{{\cal S}}
\def\CT{{\cal T}}
\def\CU{{\cal U}}
\def\CV{{\cal V}}
\def\CW{{\cal W}}
\def\CX{{\cal X}}
\def\CY{{\cal Y}}
\def\CZ{{\cal Z}}

\newcommand{\todo}[1]{{\em \small {#1}}\marginpar{$\Longleftarrow$}}
\newcommand{\labell}[1]{\label{#1}\qquad_{#1}} 
\newcommand{\bbibitem}[1]{\bibitem{#1}\marginpar{#1}}
\newcommand{\llabel}[1]{\label{#1}\marginpar{#1}}

\newcommand{\sphere}[0]{{\rm S}^3}
\newcommand{\su}[0]{{\rm SU(2)}}
\newcommand{\so}[0]{{\rm SO(4)}}
\newcommand{\bK}[0]{{\bf K}}
\newcommand{\bL}[0]{{\bf L}}
\newcommand{\bR}[0]{{\bf R}}
\newcommand{\tK}[0]{\tilde{K}}
\newcommand{\tL}[0]{\bar{L}}
\newcommand{\tR}[0]{\tilde{R}}

\newcommand{\btzm}[0]{BTZ$_{\rm M}$}
\newcommand{\ads}[1]{{\rm AdS}_{#1}}
\newcommand{\ds}[1]{{\rm dS}_{#1}}
\newcommand{\eds}[1]{{\rm EdS}_{#1}}
\newcommand{\sph}[1]{{\rm S}^{#1}}
\newcommand{\gn}[0]{G_N}
\newcommand{\SL}[0]{{\rm SL}(2,R)}
\newcommand{\cosm}[0]{R}
\newcommand{\hdim}[0]{\bar{h}}
\newcommand{\bw}[0]{\bar{w}}
\newcommand{\bz}[0]{\bar{z}}
\newcommand{\be}{\begin{equation}}
\newcommand{\ee}{\end{equation}}
\newcommand{\bea}{\begin{eqnarray}}
\newcommand{\eea}{\end{eqnarray}}
\newcommand{\pat}{\partial}
\newcommand{\lp}{\lambda_+}
\newcommand{\bx}{ {\bf x}}
\newcommand{\bk}{{\bf k}}
\newcommand{\bb}{{\bf b}}
\newcommand{\BB}{{\bf B}}
\newcommand{\tp}{\tilde{\phi}}
\newcommand{\zbar}{\bar{z}}
\newcommand{\Nbar}{\bar{N}}
\newcommand{\Ebar}{\bar{E}}
\newcommand{\patb}{\bar{\partial}}
\hyphenation{Min-kow-ski}

\def\apr{\alpha'}
\def\str{{str}}
\def\lstr{\ell_\str}
\def\gstr{g_\str}
\def\Mstr{M_\str}
\def\lpl{\ell_{pl}}
\def\Mpl{M_{pl}}
\def\varep{\varepsilon}
\def\del{\nabla}
\def\grad{\nabla}
\def\tr{\hbox{tr}}
\def\perp{\bot}
\def\half{\frac{1}{2}}
\def\p{\partial}
\def\perp{\bot}
\def\eps{\epsilon}
\newcommand{\Tr}{\mathop{\rm Tr}}


\def\NN{{\cal N}}
\def\nfour{{\cal N}=4}
\def\ntwo{{\cal N}=2}
\def\none{{\cal N}=1}
\def\nonestar{{\cal N}=1$^*$}
\def\tr{{\rm tr\ }}
\def\RR{{\cal R}}
\def\PP{{\cal P}}
\def\ZZ{{\cal Z}}

\newcommand{\bel}[1]{\be\label{#1}}
\def\al{\alpha}
\def\bt{\beta}
\def\mn{{\mu\nu}}
\newcommand{\rep}[1]{{\bf #1}}
\newcommand{\vev}[1]{\langle#1\rangle}
\newcommand{\bra}[1]{\langle#1|}
\newcommand{\ket}[1]{|#1\rangle}
\def\eref{(?FIX?)}

\renewcommand{\thepage}{\arabic{page}}
\setcounter{page}{1}

\def\Tbar{\bar{T}}
\def\pbar{\bar{\partial}}
\def\psibar{\bar{\psi}}

\def\MCA{{\mathcal A}}
\def\MCC{{\mathcal C}}
\def\MCD{{\mathcal D}}
\def\MCE{{\mathcal E}}
\def\MCG{{\mathcal G}}
\def\MCH{{\mathcal H}}
\def\MCL{{\mathcal L}}
\def\MCM{{\mathcal M}}
\def\MCN{{\mathcal N}}
\def\MCO{{\mathcal O}}
\def\MCP{{\mathcal P}}
\def\MCT{{\mathcal T}}
\def\MCV{{\mathcal V}}
\newcommand{\prm}{\textrm{ .}}
\newcommand{\crm}{\textrm{ ,}}
\newcommand{\Ave}[1]{{\left\langle{#1}\right\rangle}}

\rightline{HIP-2007-47/TH}
\vskip 1cm \centerline{\large {\bf Timelike Boundary Sine-Gordon Theory}}
\centerline{ \large {\bf and Two-Component Plasma}}
\vskip 1cm
\renewcommand{\thefootnote}{\fnsymbol{footnote}}
\centerline{{\bf Niko Jokela,$^{1}$\footnote{niko.jokela@helsinki.fi} Esko Keski-Vakkuri,$^{1,2}$\footnote{esko.keski-vakkuri@helsinki.fi}, and
Jaydeep Majumder$^{1}$\footnote{jaydeep.majumder@helsinki.fi}
}}
\vskip .5cm
\centerline{\it ${}^{1}$Helsinki Institute of Physics and ${}^{2}$Department of Physical Sciences
} \centerline{\it P.O.Box 64, FIN-00014 University of Helsinki, Finland}

\setcounter{footnote}{0}
\renewcommand{\thefootnote}{\arabic{footnote}}

\begin{abstract}
It has long been known that there is a relation between
boundary sine-Gordon theory and thermodynamics of charge neutral two-component Coulomb plasma on a unit circle. On the other
hand, recently it was found that open string worldsheet description of brane decay can be related to
a sequence of points of thermodynamic equilibrium of one-component plasma. Here we consider a different
decay process which is specifically described by the timelike boundary sine-Gordon theory. We find time evolution to
be mapped to a one-dimensional curve in the space of points of thermal equilibrium of a non-neutral
two-component Coulomb plasma. We compute the free energy of the system and find that along the curve
it is monotonously decreasing, defining a thermodynamic arrow of time.
\end{abstract}

\section{Introduction and summary}
There is an interesting proposal \cite{Sen:2002qa} to use the value of a tachyon field rolling down in its effective potential as a ``clock'', when the tachyonic system is coupled to gravity, to give a definition of time. An interesting {\emph{a priori}} unrelated question is whether the concept of time could be ``emergent'', and associated with a large $N$ limit.

Previous work on tree level string worldsheet correlation functions
in (open string) rolling tachyon background unraveled (in certain
case) $U(N)$ matrix structures \cite{nlt,Okuyama:2003jk,Gutperle:2003xf,bkkn,Shelton:2004ij,Jokela:2005ha}.
It was also understood that
correlation functions can be related to expectation values of
periodic functions in an ensemble of random $U(N)$ matrices of
varying rank \cite{bkkn, Jokela:2005ha}.
The ensemble can alternatively be interpreted as a grand canonical
ensemble of point charges on a unit circle, the Dyson gas, with different
points of thermal equilibrium labeled by different values of the
chemical potential corresponding to different instants of time \cite{bkkn, BJKM}.
Later times turn out also to be related to larger values of the average number $\bar N$ of point charges.
Worldsheet correlation functions
can then be related to thermal expectation values
of moments of the electrostatic potential in the presence of test
charges. Finally, embedding the system in a curved two-dimensional
plane, at the critical (scale invariant point) one can reinterpret
the field equations arising from the worldsheet
beta functions. In particular, time evolution equations
turn out to correspond to differential equations relating thermal
expectation values at neighboring points of thermal equilibrium, at
different chemical potential.

The $U(N)$ matrix structure was associated to a specific choice of a
rolling tachyon background, the ``half S-brane'' profile $T(X^0)=\exp
(X^0)$ on a bosonic open string theory worldsheet. Other choices
lead to other structures. For example, generalization to
superstrings leads to $U(N)\times U(M)$ type random matrix
ensembles \cite{Shelton:2004ij,Jokela:2005ha}.
However, the language of random matrices may be too
restrictive. For example, the random matrix interpretation of Dyson
gas at generic (inverse) temperature was found only relatively
recently \cite{killipnenciu}. It is easier to study what thermodynamic
ensembles correspond to different types of tachyon condensation.

This letter is a companion to \cite{BJKM} and
focuses on identifying and studying the corresponding
thermodynamic ensemble for the ``full S-brane'' rolling tachyon
profile $T(X^0)=\cosh(X^0)$. It turns out to be the grand canonical
ensemble of two-component plasma of equal but opposite charges
confined to a unit circle on a two-dimensional plane. In this case there is
no known random matrix interpretation\footnote{At least not to the
authors' knowledge.}. An interesting feature of the system is that
both components of the plasma a priori have their own chemical
potentials. Time evolution of the brane traces out a curve on the chemical
potential plane. We evaluate the Helmholtz free energy along the curve and
find it to be monotonously decreasing along the direction corresponding to later
times. This ``thermodynamic arrow of time'' is a generalization of the one
previously found in \cite{BJKM}; we also show how the latter is recovered in the
appropriate scaling limit. We are left wondering whether there could be a more general relation:
if a grand canonical statistical mechanical system with a space of points of scale invariant thermal equilibrium
has a curve along which the free energy is monotonously decreasing, the curve turns out to correspond
to the time evolution of some system in string theory.

\section{The full S-brane and the disk partition function}

We consider a rolling tachyon deformation of the open string worldsheet theory \cite{senrolling}, called the
full S-brane background. The non-trivial part of the action is given by the timelike boundary
sine-Gordon theory (TBSG) \cite{Gutperle:2003xf}

\be\label{eq:fullSpert}
 S_0 + \delta S_{\rm bdry} = -\frac{1}{2\pi} \int_{\rm disk} \partial X^0\bar \partial
 X^0�+ \lambda_0 \oint dt~ \cosh(X^0(t)) \ .
\ee
As was discussed in \cite{BJKM}, the basic quantity for
identifying the statistical mechanical system is the disk partition
function (separating
out the zero mode $X^0=x^0+X'^0$ and leaving it unintegrated),
\be\label{eq:fullSpert2}
 Z_{\rm disk}(x^0) = \int\MCD X'^0 
 e^{-S_0}e^{-\lambda_0 \oint dt~ \cosh(X^0(t))} \ .
\ee
Next one would expand the boundary perturbation in
power series, work out the contractions between
exponentials $e^{\pm X^0(t)}$ and sum up the series in the end. A subtle issue arises from
the Wick ordering of the terms in the correlators,
which require contractions of opposite types of exponentials,
and cause renormalization of the coupling constant.
Exploiting the underlying $SU(2)$ current algebra
structure \cite{Callan:1994ub} or proceeding
via fermionization \cite{Polchinski:1994my} shows that the bare coupling $\lambda_0$ gets renormalized
to\footnote{See also \cite{Kogetsu:2004te, Hasselfield:2005nv} for additional recent discussion.}
\be
  \lambda = \sin (\pi \lambda_0) \ .
\ee
After Wick ordering, all correlators are of the type
\bea\label{eq:corr}
 && G_2(t_1,\ldots ,t_{N_+};\tau_1,\ldots ,\tau_{N_-}) \equiv
 \Ave{\prod_{i=1}^{N_+}e^{X'^0(t_i)}\prod_{m=1}^{N_-}e^{-X'^0(\tau_m)}} \\
 && \ \ \ \ \ \  \ \  =
\frac{\prod_{1\leq i<j\leq N_+}|e^{it_i}-e^{it_j}|^2\prod_{1\leq m<n\leq N_-}|e^{i\tau_m}-e^{i\tau_n}|^2}
{\prod_{i=1}^{N_+}\prod_{m=1}^{N_-}|e^{it_i}-e^{i\tau_m}|^2} \prm \nonumber
\eea

The final form for the disk partition function reads \cite{senrolling}
\bea\label{eq:fulldiskpf}
 Z_{\rm disk}(x^0) && = \sum_{N_+=0}^\infty\sum_{N_-=0}^\infty\frac{(-\lambda e^{x^0})^{N_+}(-\lambda e^{-x^0})^{N_-}}
 {N_+!N_-!} \\
 \cdot &&\frac{1}{(2\pi)^{N_++N_-}} \int_{-\pi}^\pi\cdots\int_{-\pi}^\pi 
   \prod_{i=1}^{N_+} dt_i\prod_{m=1}^{N_-} d\tau_m~G(t_1,\ldots ,t_{N_+};\tau_1,\ldots ,\tau_{N_-}) \nonumber \\
 && =  \frac{1}{1+\lambda e^{x^0}}+\frac{1}{1+\lambda e^{-x^0}}-1\prm \nonumber
\eea

\subsection{The two-component plasma}

Upon analytic continuation $X^0\rightarrow iX$, the action (\ref{eq:fullSpert}) is related to the ordinary
spacelike boundary sine-Gordon theory (SBSG). More precisely, it is related to the case $\beta=2$ of the family
\be\label{eq:SG}
 S_{SG,\beta} = \frac{1}{2\pi} \int_{\rm disk} \partial X\bar \partial
 X+ \lambda_0 \oint dt~ \cos\left(\frac{\beta}{2}X(t)\right) \ .
\ee
It has been known for a long time that the tree level partition function of (\ref{eq:SG})
can be interpreted as the partition
function of a two-component plasma at inverse temperature $\beta$, confined to a unit circle in two dimensions
(see {\em e.g.} \cite{Saleur:1998hq} for a review).

However, in our case we are interested in the thermodynamic system corresponding
to the {\em timelike} theory (\ref{eq:fullSpert}) rather than the spacelike theory (\ref{eq:SG}). The key
difference is that in the spacelike theory the dependence on the zero mode $x=X-X'$ is oscillatory, and
was trivially integrated out in previous investigations.
Integration over the zero mode enforced charge neutrality, and
\cite{Fendley:1994ms} related the full partition function of
the boundary sine-Gordon theory to a grand canonical partition function of an overall charge neutral
two-component plasma, summing over the particle number of electrically neutral configurations only.
In contrast, in the timelike theory, the zero
mode dependence is important. Physically, in the case of D-brane decay,
it reflects the presence of a time-dependent source;
for example the disk partition function gives
the evolution of pressure in spacetime \cite{senrolling}. Integration of
the zero mode is associated with Fourier transform to
the canonical conjugate variable, the total energy.

The zero mode dependence necessitates a more general analysis of the associated thermodynamic system.
Overall charge neutrality is no longer enforced, hence
the grand canonical ensemble of the two-component plasma is extended to include populations
with net electric charge.
Each particle component now has its own chemical potential, to be related to the zero mode dependence.
Since there is only one
zero mode, it turns out to parameterize a one-dimensional curve in the chemical potential plane.
We begin by reviewing the details of
the grand canonical ensemble.

Consider two species of particles, $N_+$ particles carrying positive unit charge and $N_-$ particles
carrying negative unit charge, which are confined on a unit circle on a two-dimensional plane
with positions $e^{it_i}$ and $e^{i\tau_m}$. They interact via the 2-body Coulomb potential,
\be\label{eq:potential}
  V(\phi_1,\phi_2) = -\log |e^{i\phi_1}-e^{i\phi_2}|\crm
\ee
the interaction is two-dimensional while the particles
are confined to one-dimensional motion.

Let this system
be immersed in a reservoir at inverse temperature $\beta =1/T$. In the canonical partition function,
kinetic energy contributes the usual Gaussian integral over the canonical momenta. We focus on\footnote{The partition
function factorizes, and the contribution from the kinetic energy is trivial to add in if necessary, hence the
literature focuses on the potential energy contribution. Further, often the kinetic
contribution is eliminated by working in the limit of infinitely massive particles.}
the non-trivial contribution to the partition function
from the potential energy,
\bea\label{eq:canbeta}
 Z_{N_+,N_-}(\beta) & = & \frac{1}{(2\pi)^{N_++N_-}}\int_{-\pi}^\pi\cdots\int_{-\pi}^\pi\prod_{i=1}^{N_+} dt_i\prod_{m=1}^{N_-} d\tau_m \nonumber\\
  & & \cdot e^{-\beta \Big[\sum_{i<j}^{N_+}V(t_i,t_j)+\sum_{m<n}^{N_-}V(t_i,t_j)
  -\sum_i^{N_+}\sum_m^{N_-} V(t_i,\tau_m) \Big] } \\
  & = & \frac{1}{(2\pi)^{N_++N_-}} \int_{-\pi}^\pi\cdots\int_{-\pi}^\pi\prod_{i=1}^{N_+} dt_i\prod_{m=1}^{N_-} d\tau_m
 ~G_\beta (t_1,\ldots ,t_{N_+};\tau_1,\ldots ,\tau_{N_-}) \crm \nonumber
\eea
where
\be
 G_\beta (t_1,\ldots ,t_{N_+};\tau_1,\ldots ,\tau_{N_-}) = \frac{\prod_{1\le i<j\le N_+} |e^{it_i} - e^{it_j}|^{\beta}\prod_{1\le m<n\le N_-} |e^{i\tau_m} - e^{i\tau_n}|^{\beta}}
{\prod_{i=1}^{N_+}\prod_{m=1}^{N_-}|e^{it_i} - e^{i\tau_m}|^\beta} \prm
\ee
The complicated integrals can be evaluated with the help of expansion in Jack polynomials \cite{Fendley:1994ms}.
We consider first the charge neutral case $N\equiv N_+=N_-$.
The integral was found to be convergent for $\beta <1$ so at $\beta =2$ it should diverge.
On the other hand, the grand canonical partition function
\be
 Z_{\rm neutral}(z,\beta) = \sum^\infty_{N=0} \frac{z^{2N}}{N!N!} Z_{N,N}(\beta)
\ee
was found to have a distinct feature at $\beta =2$. Previous investigations \cite{Forrester} had argued that
the overall charge neutral system
undergoes a phase transition from an {\em insulating phase} for $\beta>2$ (low temperatures) to a
{\em conducting phase} for $\beta <2$ (high temperatures). In the low temperature phase, the opposite
charges tend to form dipoles and hence become insulating.
As one quantitative test \cite{Forrester}, it was argued that in the conducting phase,
the pressure of the system is non-analytic around $z=0$ in the complex fugacity plane, while in insulating
phase it is analytic around $z=0$. Ref. \cite{Fendley:1994ms} considered the system at high temperatures,
and by a saddle point analysis, they found that the
grand partition function has an essential singularity at $\beta=2$,
\be\label{essential}
 Z_{\rm neutral}(z,\beta) \sim \exp \left(z^{2/(2-\beta)} \right) \ .
\ee
However, to our knowledge a detailed understanding of the phase structure of the system is still lacking.

Let us then consider the general case $N_+\neq N_-$. The canonical partition function (following
\cite{Fendley:1994ms}) is
\be\label{e6} Z_{N_+,N_-}(\beta) =  Z_{N_+}(\beta) Z_{N_-} (\beta)
\left\{\sum_{{\lambda \atop \ell(\lambda)\le {\rm min}(N_+,N_-)}}
N_\lambda(N_+)N_\lambda(N_-) \right\}\crm
\ee
where
\be\label{eq:jellium}
 Z_N(\beta) = \frac{\Gamma(1 +\frac{\beta N}{2})}{\left(\Gamma(1+\frac{\beta}{2})\right)^N}
\ee
is the canonical partition function for a single charge Coulomb gas on a circle (the Dyson gas), and
\be\label{e7}
 N_\lambda(N) = 
 \prod_{s\in\lambda}\left(\frac{j-1+\frac{\beta}{2}(N-i+1)}
 {j+\frac{\beta}{2}(N-i)}\right)\prm
\ee
The sum in (\ref{e6}) is over partitions with an upper bound on length
and the product in (\ref{e7}) is over
the cells of a partition.
Particular cases are $Z_{N_+,0}(\beta )=Z_{N_+}(\beta ), Z_{0,N_-}(\beta )=Z_{N_-}(\beta )$.
Consider then the special value $\beta=2$, where by the example (\ref{essential})
we would expect the partition function
(\ref{e6}) to diverge. However, guided by the disk partition function (\ref{eq:fulldiskpf}),
one can adopt a prescription to regulate the integrals, the details are given in Appendix.
The regularized partition function simplifies miraculously to
\be\label{Zregu}
 Z_{N_+,N_-}(\beta =2) = N_+!N_-!(\delta_{N_+,0}+\delta_{N_-,0}-\delta_{N_++N_-,0}) \prm
\ee
Next we allow particle exchange with the reservoir and move to
the grand canonical ensemble. Unlike in the previous investigations, we let the opposite
charged particles have independent populations with chemical potentials $\mu_\pm$, so
the grand partition function of the system is
\bea\label{eq:Z2Galfa}
  Z_{2G}(z_+,z_-,\beta) = \sum^\infty_{N_+=0}\sum^\infty_{N_-=0} \frac{z_+^{N_+}z_-^{N_-}}{N_+!N_-!}
    Z_{N_+,N_-}(\beta) \crm
\eea
With the regulated expression  (\ref{Zregu}), at inverse temperature $\beta=2$ (\ref{eq:Z2Galfa})
simplifies to
\be\label{eq:Z2G}
 Z_{2G}(z_+,z_-,\beta =2) = \frac{1}{1-z_+}+\frac{1}{1-z_-}-1 \crm
\ee
resembling the disk partition function (\ref{eq:fulldiskpf}). Note that away from the poles at $z_\pm=1$
the partition function is regular, in contrast to (\ref{essential}).

As in the previous investigation \cite{BJKM}, relating
the grand partition function of the thermodynamic system
to the worldsheet disk partition function will require analytic continuation,
to map chemical potential to time (the explicit zero mode of (\ref{eq:fulldiskpf})).
But now there is a new interesting twist: the two components of the plasma come with two chemical
potentials but there is only one physical time\footnote{We thank S. Brodsky for this remark.}.
So one must restrict to a subset of points of chemical equilibrium, corresponding
to a one-dimensional curve in the two-dimensional chemical potential (or fugacity) plane.
Following \cite{BJKM}, we start from (\ref{eq:Z2G}), then analytically continue to negative fugacities, which
we identify with real values of time $x^0$:
\be\label{eq:analyt}
  z_{\pm} \rightarrow -z_{\pm} = -\lambda e^{\pm x^0} \ ,
\ee
which defines a one-dimensional curve $z_- = \lambda^2 / z_+$ in the fugacity plane.

In this way, the grand partition function (\ref{eq:Z2G}) is mapped to the disk partition function
(\ref{eq:fulldiskpf}). Note that we need to restrict to values $|z_\pm| <1$ which correspond to
\be\label{life}
   |x^0| < \frac{\Delta x^0}{2} \equiv -\log \lambda \prm
\ee
It is interesting that the real time range $\Delta x^0$,
corresponding to the allowed values of the fugacities, has a
very natural physical interpretation:
the lifetime of the brane\footnote{{\em I.e.}, the time interval during which the unstable
brane first forms and then decays, see {\em e.g.} \cite{Gaiotto:2003rm}.}.

\subsection{Thermodynamic arrow of time}

In the case of the half S-brane,
it was found that $\bar N$, the average number of charges in the Dyson gas, could be used as the
time variable, and the free energy of the system was a monotonically decreasing function of $\bar N$,
giving an arrow of time. This analysis was done at the temperature $\beta=2$ corresponding to
the exactly marginal rolling tachyon deformation \cite{BJKM}. Here we extend the calculations for
the full S-brane. Extension to superstrings will be considered in \cite{jimmyniko}.

The average particle numbers for positive and negative charges are
\be
 \bar N_\pm \equiv \frac{1}{Z_{2G}}\sum_{N_+=0}^\infty\sum_{N_-=0}^\infty N_\pm \frac{z_+^{N_+}z_-^{N_-}}{N_+!N_-!}Z_{N_+,N_-} = \frac{1-z_\mp}{1-\lambda^2}\frac{z_\pm}{1-z_\pm} \prm
\ee
Restricting to the curve $z_+ z_- = \lambda^2$, the numbers $\bar N_\pm$ are related by
\be
 \bar N_+ \bar N_- = \frac{\lambda^2}{(1-\lambda^2)^2}\prm
\ee
To specify the thermodynamics of the system onto the curve, we solve the constraint and express
the grand partition function
in terms of $\mu_+$, 
\be
Z_{2G} = \frac{(1-\lambda^2)e^{\beta\mu_+}}{(1-e^{\beta\mu_+})(e^{\beta\mu_+}-\lambda^2)}\crm
\ee
Then, on the curve the Legendre transform to the Helmholtz free energy is given by
\be
 A_{\rm curve}(\bar N_+) = -\beta^{-1}\log Z_{2G}+\beta^{-1}\mu_+\frac{\partial}{\partial\mu_+}\log Z_{2G} \ .
\ee
At $\beta =2$, the free energy along the curve becomes
\bea\label{helmholz}
 && A_{\rm curve}(\bar N_+) =
 \half\log\bar N_+ -\half\log(1+(1-\lambda^2)\bar N_+)-\half\bar N_+\log(1+(1-\lambda^2)\bar N_+) \nonumber\\
 && \ \ \ \ \  +\half\frac{\lambda^2}{(1-\lambda^2)^2\bar N_+}\log(1+(1-\lambda^2)\bar N_+)
    -\half\log(\lambda^2+(1-\lambda^2)\bar N_+) + \log(1-\lambda^2) \nonumber\\
 && \ \ \ \ \  +\half\bar N_+\log(\lambda^2+(1-\lambda^2)\bar N_+)
                   -\half\frac{\lambda^2}{(1-\lambda^2)^2\bar N_+}
                   \log(\lambda^2+(1-\lambda^2)\bar N_+)  \prm 
\eea
It is straightforward to check
that the free energy $A_{\rm curve}$ is indeed monotonously decreasing as a function of
the curve parameter $\bar N_+$. So it can be interpreted
as a thermodynamic arrow of time, in  the sense of \cite{BJKM}.

\subsection{Dyson gas as a scaling limit of the two-component plasma}
Recall that to get half S-brane, we are shifting the origin of the time coordinate as follows
\be
 \lambda_0 \int dt \cosh(X^0(t)+C) =  \half\lambda_0\int dt (e^{X^0+C}+e^{-X^0-C})
 \longrightarrow
 \lambda\int dt e^{X^0}\crm
\ee
where the limit
\bea\label{limit}
 \lambda_0 \to  0
  \ , \ C \to  \infty
\ , \ \lambda_0  e^C  \equiv 2\lambda  = { \ \textrm{fixed}}
\eea
was applied.
This limit is also realized in Dyson gas.
Recall the partition function (\ref{eq:Z2G}).
First we implement the shift $x^0 \to x^0 + C$:
\bea
 z_\pm & = & \hat\lambda e^{\pm (x^0+C)}\prm
\eea
and use a different notation for the renormalized coupling,
$\hat\lambda = \sin(\pi\lambda_0)$.
Now we apply the limit (\ref{limit})
for the two-component plasma,
\bea
 \hat \lambda  & \to & \pi \lambda_0 \to 0 \\
 z_- = \hat \lambda e^{-x^0-C} & \to & 0 \\
 z_+ = \hat \lambda e^{x^0+C} & \to & \pi \lambda_0e^C e^{x^0}\equiv 2\pi\lambda e^{x^0} = z\crm
\eea
so the population of negative charges goes to zero while the positive charges remain.
In more detail,
\bea
 \bar N_+ \to \frac{z}{1-z} \  ;  \  \delta N_+ \to \frac{1}{\sqrt z}
\eea
while
\bea
 \bar N_- \to 0  \  ;  \ \delta N_- \to 0\prm
\eea
Also the Helmholtz free energy (\ref{helmholz}) reduces to the previous result obtained
in the case of the half S-brane
\be
 A_{\rm curve}(\bar N_+) \to -\half[ (\bar N_+ +1)\log(\bar N_+ +1)-\bar N_+\log \bar N_+ ]\prm
\ee
In particular, the grand partition function reduces to
\bea
 Z_{2G}\to \frac{1}{1-z} = Z_G\crm
\eea
the partition function of the one-component plasma (Dyson gas).

\bigskip

\noindent
{\bf \large Acknowledgments}

\bigskip

We thank V. Balasubramanian, M. J\"arvinen, K. Kyt\"ol\"a, and A. Naqvi for useful discussions.
N.J. and  J.M. thank the University of
Pennsylvania for hospitality.
N.J. has been in part supported by the Magnus Ehrnrooth foundation.
E.K-V. and J.M. were in part supported by
the Academy of Finland.  This work was
also partially supported by the EU 6th Framework Marie Curie
Research and Training network ``UniverseNet'' (MRTN-CT-2006-035863).

\appendix
\section{Partition Function at $\beta=2$}\label{app:fullpart}
The purpose of this appendix is to fill in the gaps between (\ref{eq:Z2Galfa}) and (\ref{eq:Z2G}).
Let us begin from (\ref{eq:Z2Galfa}). It is straightforward to  manipulate
\bea
 \sum_{N_+=0}^\infty\sum_{N_-=0}^\infty \frac{(z_+)^{N_+}(z_-)^{N_-}}{N_+!N_-!}Z_{N_+,N_-} 
   & = & \sum_{N_+=1}^\infty\sum_{N_-=1}^\infty
   \frac{(z_+)^{N_+}(z_-)^{N_-}}{N_+!N_-!}Z_{N_+,N_-}\nonumber\\
   & & + \frac{1}{1-z_+} + \frac{1}{1-z_-} -1 \crm
\eea
where
\bea
 Z_{N_+,N_-}
 & = & \frac{1}{(2\pi)^{N_++N_-}}\int_{-\pi}^\pi\cdots\int_{-\pi}^\pi dt_1\cdots dt_{N_+}d\tau_1\cdots d\tau_{N_-} \nonumber\\
 & & \cdot \frac{\prod_{1\leq i<j\leq N_+}|e^{it_i}-e^{it_j}|^2\prod_{1\leq m<n\leq N_-}|e^{i\tau_m}-e^{i\tau_n}|^2}{\prod_{i=1}^{N_+}\prod_{m=1}^{N_-}|e^{it_i}-e^{i\tau_m}|^2}\label{eq:Zpm}\prm
\eea

The only non-trivial part is to show the vanishing of $Z_{N_+,N_-}$ with $N_+,N_-\ne 0$.
To achieve this, define $z_i = e^{it_i}$ and $w_m = e^{i\tau_m}$. Then, (\ref{eq:Zpm}) becomes
\bea
 Z_{N_+,N_-}
  & = & \frac{(-i)^{N_++N_-}}{(2\pi)^{N_++N_-}}\oint_\MCC\cdots\oint_\MCC \frac{dz_1}{z_1}\cdots \frac{dz_{N_+}}{z_{N_+}} \frac{dw_1}{w_1}\cdots \frac{dw_{N_-}}{w_{N_-}}\nonumber\\
 & & \cdot\frac{\prod_{1\leq i<j\leq N_+}(z_i-z_j)(\frac{1}{z_i}-\frac{1}{z_j})\prod_{1\leq m<n\leq N_-}(w_m-w_n)(\frac{1}{w_m}-\frac{1}{w_n})}{\prod_{i=1}^{N_+}\prod_{m=1}^{N_-}(z_i-w_m)(\frac{1}{z_i}-\frac{1}{w_m})} \\
 & = & \frac{(-i)^{N_++N_-}}{(2\pi)^{N_++N_-}}\oint_\MCC\cdots\oint_\MCC dz_1\cdots dz_{N_+}dw_1\cdots dw_{N_-}\nonumber\\
 &  & \cdot\left(\frac{1}{z_1\cdots z_{N_+}}\right)^{N_+-N_-}\left(\frac{1}{w_1\cdots w_{N_-}}\right)^{N_--N_+}\nonumber\\
 &  &  \cdot\frac{\prod_{1\leq i<j\leq N_+}\left[-(z_i-z_j)^2\right]\prod_{1\leq m<n\leq N_-}\left[-(w_m-w_n)^2\right]}{\prod_{i=1}^{N_+}\prod_{m=1}^{N_-}\left[-(z_i-w_m)^2\right]}\crm
\eea
where $\MCC$ is the unit circle (integrated counterclockwise). 

Due to the double poles, one needs to properly define a regularization prescription.
We use the principal value prescription, defined as
\be
 \frac{1}{(z_i-w_m)^2} \to \half\frac{1}{(z_i-C w_m)^2} +\half\frac{1}{(C z_i-w_m)^2} \equiv P\frac{1}{(z_i-w_m)^2} \ \crm \ \ C=1+\eps\prm
\ee

Let us now integrate over one variable, and without any loss of generality we can choose $z_1$, assuming $N_+\leq N_-$\footnote{If $N_+>N_-$ we will choose $w_1$.}. Now
\bea
 Z_{N_+,N_-} &\sim & \oint_\MCC dz_1\frac{1}{z_1^{N_+-N_-}}\prod_{j=2}^{N_+}\left[-(z_1-z_j)^2\right]\cdot\prod_{m=1}^{N_-}\left[-P\frac{1}{(z_1-w_m)^2}\right]\\
 & \equiv & \oint_\MCC dz_1 g(z_1)\equiv J\prm
\eea
At infinity
\be
 g(z_1)\sim\frac{1}{z_1^{N_+-N_-}}z_1^{2(N_+-1)}z_1^{-2N_-}=\frac{1}{z_1^2}\frac{1}{z_1^{N_--N_+}}\crm
\ee hence there is no residue at $z_1=\infty$. At zero there is no residue as well, since
\be
 g(z_1)\sim z_1^{N_--N_+}\prm
\ee
We thus conclude that all the residues of $g$ are lying on the integration contour. 
The principal value prescription gives one half of each residue contribution,
\be
 Z_{N_+>0,N_->0}\sim J = \pi i\sum{\rm{Res}}\ g\crm
\ee
where
\be
\sum{\rm{Res}}\ g = 0
\ee
as the sum of the residues of any rational function is zero.

\end{document}